\documentclass[aps,prd,showpacs]{revtex4}
\usepackage{graphicx}
\usepackage{epsfig}
\usepackage{psfig}
\usepackage{revsymb}
\begin{document}
\renewcommand{\thesection}{\arabic{section}}
\renewcommand{\thesubsection}{\arabic{subsection}}

\title{On The Problem with Only Zeroth Landau Level Occupancy of Electrons 
in Strongly Magnetized Neutron Star Matter in $\beta$-equilibrium 
Condition- The Role of Magnetic Dipole Moments}

\author{
 Sutapa Ghosh$^{a)}$, Soma Mandal$^{a)}$ and Somenath Chakrabarty$^{b)}$
}
\affiliation{
$^{a)}$Department of Physics, University of Kalyani, Kalyani 741 235,
India\\
$^{b)}$Department of Physics, Visva-Bharati, Santiniketan 731 235,
India, E-mail:somenath@vbphysics.net.in
}
\pacs{97.60.Jd, 97.60.-s, 75.25.+z} 

\begin{abstract}
It is shown explicitly that if only the zeroth Landau level is occupied by
electrons in  strongly magnetized neutron star matter in
$\beta$-equilibrium, then both the proton and neutron matter
sectors become spin symmetric. Whereas, the study of Pauli 
para-magnetism of neutron matter sector shows that such a scenario is 
physically impossible. It is also shown that in dense stellar hadronic matter 
in $\beta$-equilibrium in presence of strong quantizing magnetic field, 
with $\sigma-\omega-\rho$ exchange type mean field interaction and with the 
inclusion of magnetic dipole moments does not allow electrons to occupy 
only the zeroth Landau level.
\end{abstract}
\maketitle

The study of dense stellar matter in presence of strong magnetic 
field, which is one of the oldest subject of physics, has gotten a new life 
after the discovery of a few magnetars, which are believed to be strongly 
magnetized young neutron stars \cite{R1,R2,R3,R4}. The surface magnetic field 
of these exotic objects are observed to be $\geq 10^{15}$G. It is then quite 
possible that the field at the 
core region can go up to $10^{18}$G. Although the origin of such strong 
magnetic field is not known till today, the studies on the 
effect of such strong magnetic field on various physical properties and 
processes in neutron stars particularly at the core region
are extremely interesting and important fields of 
research in nuclear astrophysics as well as in high energy astrophysics 
\cite{R5,R6,R7,R8,SC1,SC2}. Not only that, the general relativistic studies of such 
objects have also got some importance \cite{R9,R10,R11}. The
studies of physical  properties of magnetically deformed stellar objects
with general relativity are found to be extremely interesting.

An extensive studies have already been done on the equation of state of
strongly magnetized neutron stars \cite{R5,R6,SC1,SC2}. The effect of strong magnetic
field on the physical processes, e.g., the weak reactions and decays
have also been studied in neutron star matter \cite{R7,R8}. As a consequence 
the cooling rate of neutron stars by neutrino emission also change 
significantly in presence of strong magnetic fields. It is also found that 
the effect of strong magnetic field on all these physical aspects of neutron stars become
most prominent if the charged particles (e.g., electron, proton, etc.) occupy
only their zeroth Landau levels. 

Now it is known that the maximum value of Landau quantum number occupied
by electrons at zero temperature is given by
\begin{equation}
n_{\rm{max}}^{(e)}=\frac{[(\mu_e^2-m_e^2)}{2eB}]
\end{equation}
where $[~]$ indicates
the integer part only. Obviously, this maximum value is a function of
both magnetic field strength and density of stellar matter through
$\mu_e$, the electron chemical potential, $m_e$ and $e$ are the corresponding rest mass and the magnitude of charge. Here we assume the gauge $A^\mu\equiv (0,0,xB,0)$, so that the magnetic field $B$ is along x-direction and is constant. Then it is very easy to show, that for electrons, this quantity
becomes zero if $B\sim 10^{17}$G in dense stellar matter of
astrophysical interest. 

The aim of this article is to show that if the
magnetic field in a neutron star is high enough so that the electrons occupy 
only their zeroth Landau level,
then there will be a serious problem on the existence of predicted spin
symmetric neutron and proton components in 
dense neutron star matter. We shall also show that the same 
conclusion is also true for interacting hadronic matter with non-zero
magnetic dipole moments of the constituents, including the anomalous magnetic moment of neutron. In fig.(1) we have shown how 
$n_{\rm{max}}^{(e)}=0$ value for electrons depends on the strength of magnetic field 
and density of stellar matter. The curve in this figure indicates
$n_{\rm{max}}^{(e)}=1$ for electrons as a function of magnetic field
strength and matter density. The space above the line is for 
$n_{\rm{max}}^{(e)}=0$, whereas below this line, all possible non-zero allowed values.
The curve therefore
separates the region where only zeroth Landau level is populated
from the region where other non-zero Landau levels are occupied by electrons. 

To sketch the curve as shown in fig.(1), we set $n_{\rm{max}}^{(e)}=1$
then it is very easy to show that 
\begin{equation}
\mu_e=m_e(1+2h_f)^{1/2}
\end{equation}
where $h_f=B/B_c^{(e)}$ and $B_c^{(e)}=4.41\times 10^{13}$G, the quantum 
mechanical limit of magnetic field strength for electrons (this is the
typical strength of magnetic field at which the value of cyclotron quantum
for
electron exceeds the corresponding rest mass
energy). 
Hence we can write down for 
the electron density assuming that they are moving freely within the system 
\begin{equation}
n_e=\frac{eB}{2\pi^2} \sum_{n=0}^{n_{\rm{max}}^{(e)}(=0)}(2-\delta_{n0})
(\mu_e^2-m_e^2-2n e
B)^{1/2} = n_p {~~~~\rm{(from ~charge ~ neutrality ~ condition)}}
\end{equation}
Again the proton density is given by (here also for the sake of simplicity we 
have not considered interaction)
\begin{equation}
n_p=\frac{eB}{2\pi^2} \sum_{n=0}^{n_{\rm{max}}^{(p)}}(2-\delta_{n0}) 
(\mu_p^2-m_p^2-2n e B)^{1/2}
\end{equation}
where $\mu_p$ and $m_p$ are respectively the chemical potential and the mass
of proton.
Now substituting eqn.(4) into eqn.(3), we can solve numerically for $\mu_p$  
for a given $B$. Then assuming  $\beta$-equilibrium condition, 
we have $\mu_n=\mu_p +\mu_e$, the neutron chemical potential.
We assume that neutrinos / anti-neutrinos are non-degenerate, i.e.,
leave the system as soon as they are created. Then
using the relations
\begin{equation}
n_n=\frac{(\mu_n^2-m_n^2)^{3/2}}{3\pi^2}
\end{equation}
for neutron density. We have $n_B=n_p+n_n$, the total baryon number density.

Now in neutron stars the dynamic nature of chemical equilibrium among the
constituents is mainly controlled by the following weak interaction processes 
(URCA and modified URCA processes \cite{R12}). These are also the most important
cooling processes by neutrino / anti-neutrino emissions and are given by
\begin{eqnarray}
n &\rightarrow & p+ e^- +\bar \nu_e \nonumber \\
p+ e^- & \rightarrow & n + \nu_e
\end{eqnarray}
Now from the solution of Dirac equation for electrons in presence of a
quantizing magnetic field it can be shown that
when only the zeroth Landau level is 
occupied by electrons, all the
electrons Will be aligned in the opposite direction to
the magnetic field.
Then we can write down the following relations from the condition 
of dynamical chemical equilibrium
\begin{equation}
\mu_{p\uparrow}=\mu_{p\downarrow}, ~~~\mu_{n\uparrow}=\mu_{n\downarrow}
\end{equation}
with
\begin{eqnarray}
\mu_{n\uparrow}&=&\mu_{p\uparrow}+\mu_{e\downarrow}
\nonumber \cr
\mu_{n\downarrow}&=&\mu_{p\downarrow}+\mu_{e\downarrow}
=\mu_{p\uparrow}+\mu_{e\downarrow}
\end{eqnarray}
Here we have indicated the direction of polarization of electrons by
$\downarrow$. 
Which simply means that
both the proton and neutron matter sectors are exactly spin symmetric. 
Although neutrinos / anti-neutrinos are assumed to be non-degenerate, they 
must carry the spin signature of the processes in which they are produced.
This scenario of neutrino / anti-neutrino emissions are almost identical with 
the famous experiment by C.S. Wu using polarized Co$^{60}$. The only
difference is that here instead of polarized nucleons, we have polarized
electrons, which are either absorbed or emitted by the nucleons.
To show, that the spin symmetric picture of both neutron and proton
sector in a strongly magnetized neutron star is physically impossible
we investigate the Pauli para-magnetism for neutron component.
We are
not considering the proton matter, since they may occupy some of their non-zero 
Landau levels and as a consequence the quantum dia-magnetism of proton matter 
may dominate over the Pauli para-magnetism. Now it is very easy to show
that the total neutron number density in the relativistic region is
given by
\begin{equation}
n_n=\frac{[2\mu_n^*B(2\mu_n^*B+2m_n)]^{3/2}}{6\pi^2}(2x_n^{3/2} -1)
\end{equation}
where $\mu_n^*$ is the magnitude of neutron anomalous magnetic dipole moment and
\begin{equation}
x_n=\frac{\varepsilon_n^2(B)-m_n^2}{2\mu_n^*B(2\mu_n^*B+2m_n)},
\end{equation}
$\varepsilon_n(B)$ is the single particle energy for neutron in presence of the
magnetic field $B$, which is given by 
\begin{equation}
\varepsilon_n(B)=\{m_n^2+x_n[R_{3/2}]^{-2/3}(\varepsilon_0^2-m_n^2)\}^{1/2}
\end{equation}
where $\varepsilon_0$ is the value of $\varepsilon_n(B)$ in absence of
magnetic field.  Now we define
\begin{equation}
R_{3/2}=\frac{1}{2} [2x_n^{3/2}-1]
\end{equation}
then the density of up-spin neutrons is given by
\begin{equation}
n_n^\uparrow (B) =\frac{n_n}{2} x_n^{3/2} [R_{3/2}(x_n)]^{-1}
\end{equation}
and similarly for the down-spin case we have
\begin{equation}
n_n^\downarrow (B)= n_n^\uparrow [1- x_n^{-3/2}]
\end{equation}
Now for $x=\infty$, which corresponds to field free case, then we have $n_n^\uparrow
=n_n^\downarrow=n_n/2$, which indicates that both  up and down spin
neutron states are equi-probable. On the other hand for $x\rightarrow 1$, we
have $n_n^\uparrow(B)=n_n$ and $n_n^\downarrow(B)=0$,
where this limiting condition indicates the complete saturation,  the
corresponding strength of magnetic field is given by 
\begin{equation}
B_s=\frac{1}{2\mu_n^*} [\{(6\pi^2n_n)^{2/3} +m_n^2\}^{1/2}-m_n]
\end{equation}
and the single particle energy for neutron in the saturation limit is given by
\begin{equation}
\varepsilon^{(s)}(B=B_s)=[2^{2/3}\varepsilon_0^2-(2^{2/3}-1)m_n^2]^{1/2}
\end{equation}
In fig.(2) we have plotted $n_n^\uparrow$ and  $n_n^\downarrow$ for various 
values of baryon number densities having astrophysical relevance. We have 
obtained the corresponding magnetic field strength from fig.(1). In these 
studies, since we are interested only on those electrons occupying the zeroth
Landau level, we have taken $B$ from fig.(1) slightly above the
$n_{\rm{max}}^{(e)}=1$ line ($=2.5 \times B$ of $n_{\rm{max}}^{(e)}=1$ line). 
This figure shows that within the range of densities
and magnetic fields of astrophysical relevance, neutron matter sector 
can not be spin symmetric ($n_n^\uparrow \neq n_n^\downarrow$).

We shall now go over to interacting picture of hadronic matter. We
assume a $\sigma-\omega-\rho$ exchange type mean field interaction 
scenario. Then the chemical potentials are given by
\begin{eqnarray}
\mu_p&=&\epsilon_f^p+g_\omega \omega_0 +\frac{1}{2}g_\rho
\rho_3^0\nonumber \\
\mu_n&=&\epsilon_f^n+g_\omega \omega_0 -\frac{1}{2}g_\rho
\rho_3^0\nonumber \\
\mu_e&=&\epsilon_f^e
\end{eqnarray}
where $\omega_0$ and $\rho_3^0$ are the mean values for vector and
the third component of iso-vector fields respectively, 
$g_i$'s are the corresponding 
coupling constants ($i=\sigma$, $\omega$ and $\rho$). 
Assuming the non-zero values for the magnetic moment 
for the constituents, the Fermi energies
$\epsilon_f^i$'s ($i=p,n$ and $e$) are then given by
\begin{eqnarray}
\epsilon_f^p&=& \left (p_f^{p^2} + ((m^{*^2}+2eB\nu)^{1/2}+\Delta_p
s_p B)^2\right )^{1/2} \nonumber \\
\epsilon_f^n&=& \left (p_f^{n^2} + (m^*+\Delta_n
s_n B )^2\right )^{1/2} \nonumber \\
\epsilon_f^e&=& \left (p_f^{e^2} + ((m^2+2eB\nu)^{1/2}+\Delta_e
s_e B)^2 \right )^{1/2} 
\end{eqnarray}
where $m*=m_n-g_\sigma \sigma_0$, the nucleon effective mass,
$\Delta_i$'s are related to the magnetic moments of the
components, including the neutron part, $s_i=+1$ and $-1$ for spin  up and spin down cases respectively. 
If we neglect the anomalous magnetic moments, we get back the results as 
presented for the non-interacting case, with bare nucleon mass replaced by the
effective mass. In that case it is again very easy to show that the spin symmetric 
neutron-proton matter can not exist inside a strongly magnetized neutron star.
On the other hand in presence of (anomalous) magnetic 
moments, either we must have
\begin{equation}
\mu_{p\uparrow} \neq \mu_{p\downarrow}, ~~~\mu_{n\uparrow}\neq \mu_{n\downarrow}
\end{equation}
instead of eqn.(7), or $\Delta_p=\Delta_n=0$ to satisfy eqn.(7). It is
obvious that this inequality is not consistent with the $\beta$-equilibrium
condition. On the other hand the assumption of zero (anomalous) magnetic
moments is against the established quantum mechanical results. Hence in
this case also the possibility of spin symmetric neutron matter or
proton matter is questionable.

We therefore conclude that if the magnetic field strength inside a neutron
star is such that the electrons 
occupy only the zeroth Landau level and are in $\beta$-equilibrium with
neutron and proton components,
then the necessary conditions of spin symmetry are not satisfied.
Alternatively, we may conclude that the magnetic field strength can not go
beyond $10^{17}$G, so that electrons occupy only their zeroth Landau level.

\newpage
\begin{figure}[ht]
\psfig{figure=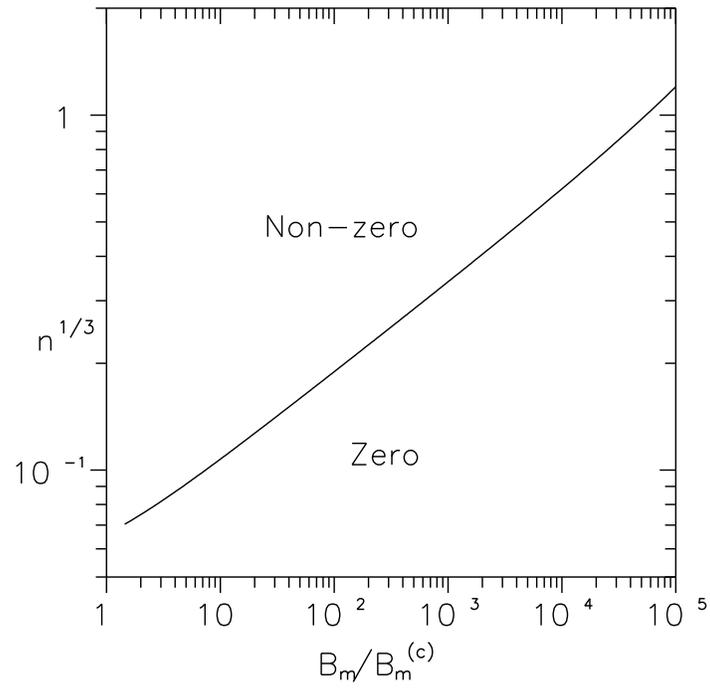,height=0.5\linewidth}
\caption{
The curve separating  only zero Landau level occupied region and
non-zero Landau level occupied region for electrons. 
}
\end{figure}
\begin{figure}[ht]
\psfig{figure=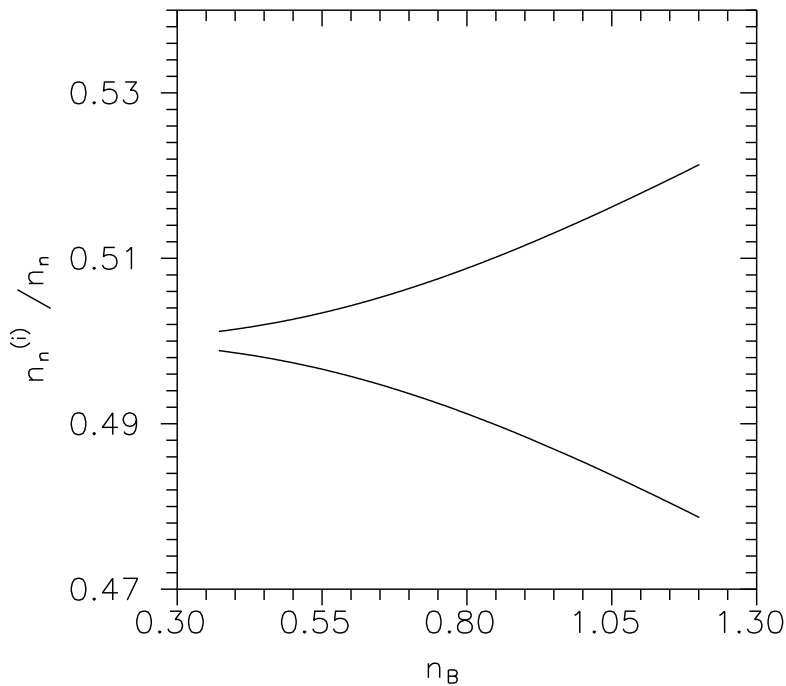,height=0.5\linewidth}
\caption{
Variation of neutron number density $n_n^{(i)}$ (expressed in terms of
$n_n$) for a spin either parallel or anti-parallel to the external
magnetic field $B$, with the baryon number density $n_B$ (expressed in
fm$^{-1}$). The upper curve is for
$n_n^\uparrow$ and lower one for $n_n^\downarrow$. In this case
electrons are occupying only their zeroth Landau level.
}
\end{figure}

\end{document}